\documentclass[12pt]{iopart}

\usepackage{amsmath,amsfonts}
\begin{document}
	
	\title[Cosmography of the f(R,T) gravity theory]{Cosmography of the \textit{\textbf{f(R,T)}} gravity theory}
	
	\author{I.S. Farias$^a$ and P.H.R.S. Moraes$^{a,b}$}
	
	\address{$^a$Instituto Tecnológico de Aeronáutica, Departamento de Física, Centro Técnico Aeroespacial, 12228-900, São José dos Campos, São Paulo, Brazil}
	\address{$^b$Universidade de São Paulo, Instituto de Astronomia, Geofísica e Ciências Atmosféricas, Rua do Matão 1226, Cidade Universitária, 05508-090, São Paulo-SP, Brazil}
	\ead{i.s.farias@outlook.com, moraes.phrs@gmail.com}
	\vspace{10pt}
	\begin{indented}
		\item[]July 2021
	\end{indented}
	
	\begin{abstract}
		Currently, in order to explain the accelerated expansion phase of the universe, several alternative approaches have been proposed, among which the most common are dark energy models and alternative theories of gravity. Although these approaches rest on very different physical aspects, it has been shown that both are in agreement with the data in the current status of cosmological observations, thus leading to an enormous degeneration between these models. So until evidences of higher experimental accuracy are available, more conservative model independent approaches are a useful tool for breaking this degenerated cosmological models picture. Cosmography as a kinematic study of the universe is the most popular candidate on this regard. Here we show how to construct the cosmographic equations for the $f(R,T)$ theory of gravity within a conservative scenario of this theory, where $R$ is the Ricci curvature scalar and $T$ is the trace of the energy-moment tensor. Such equations relate $f(R,T)$ and its derivatives at the current time $t_0$ to the cosmographic parameters $q_0$, $j_0$ and $s_0$. In addition, we show how these equations can be written within different dark energy scenarios, thus helping to discriminate between them. We also show how different $f(R,T)$ gravity models can be constrained using these cosmographic equations.
	\end{abstract}
	
	\submitto{\JCAP}
	%\maketitle

	\section{Introduction}\label{sec:intro}
	The universe is currently passing though a phase of accelerated expansion. This has been firstly brought to the scientific community attention through observations of type Ia supernovae \cite{riess/1998,perlmutter/1999}, whose brightness was lower than theoretically expected. The low brightness observations are an indication that the objects were more distant than expected, thus the Hubble expansion should be more rapid. Indeed, until such an astronomical milestone, it was believed that the universe was expanding in a decelerated form. This was intuitive in a universe dominated by gravity, which, as an attractive force should slow down the expansion. 
	
	It is worth to remark that the universe expansion started accelerating ``recently''. Some constraints have been put to the transition (from a decelerated to an accelerated expansion) redshift, $z_{t}$. From $H(z)$ observational data, with $H$ being the Hubble parameter, it was found that $z_t=0.72\pm0.05$ and $z_t=0.84\pm0.03$, respectively for $H_0=68\pm2.8$km/s/Mpc and $H_0=73.24\pm1.74$km/s/Mpc \cite{farooq/2017}. Combining 20 gamma-ray bursts and 157 type Ia supernovae, in \cite{wang/2006} it was obtained that $z_t=0.61^{+0.06}_{-0.05}$. Kinematic constraints were put to $z_t$ in \cite{cunha/2009} from the following parametrization of the deceleration parameter $q(z)=q_0+q_1z$, with $q_0$ and $q_1$ constants to be constrained by supernovae data. The result was $z_t=0.49^{+0.14}_{-0.07}$.
	
	Before started accelerating its expansion, the universe dynamics was dominated by matter and radiation \cite{ryden/2003}, as redshift increases. Right after the Big-Bang, remarkably the universe passed through its first phase of accelerated expansion: the inflationary era \cite{guth/1981}. Although the first accelerated stage of the universe expansion lasted significantly less than the current one, its consequences were pretty relevant even for the current universe \cite{popa/2011}-\cite{guzzetti/2016}.
	
	Due to the counter-intuitive feature of the accelerated expansion of the universe, to describe such a phenomenon is still a challenge. The cosmological constant $\Lambda$ in the Einstein's field equations of General Relativity,
	
	\begin{equation}\label{i2}
		G_{\mu\nu}=\frac{8\pi G_N}{c^4}T_{\mu\nu}-\Lambda g_{\mu\nu},
	\end{equation}
	satisfactorily plays this role. In (\ref{i2}), $G_{\mu\nu}$ is the Einstein tensor, $G_N$ is the newtonian gravitational constant, $c$ is the speed of light, $T_{\mu\nu}$ is the energy-momentum tensor and $g_{\mu\nu}$ is the metric. 
	
	However there is a trouble when physically interpreting $\Lambda$. Due to its repulsive feature, $\Lambda$ is related the vacuum quantum energy, with equation of state (EoS) $p=-\rho$, with $p$ being the pressure and $\rho$ the matter-energy density. The vacuum quantum energy can be calculated via Particle Physics and what is found  \cite{weinberg/1989} is $\sim120$ orders of magnitude different from the value of $\Lambda$ needed to fit cosmological observations \cite{riess/1998,perlmutter/1999}.
	
	This has led a number of theoretical physicists to develop alternatives to describe the recent cosmic acceleration with no need for $\Lambda$. These alternatives can be cast in the Extended Gravity Theories, in the sense they {\it extend} General Theory of  Relativity by incorporating new degrees of freedom, that can result, for example, in the important cosmic acceleration observable.
	
	Anyhow it is important to mention that over the years, not only the recent  cosmic acceleration was subjected to the Extended Gravity Theories application, but also inflation, dark matter, black holes, neutron stars, gravastars, gravitational waves and many other topics, as it can also be checked in the references below.
	
	There is a large number of Extended Gravity Theories nowadays. One can see, for instance, References \cite{nesseris/2006}-\cite{capozziello/1998}, and for some reviews, one can check \cite{capozziello/2011,capozziello/2008}. Most of them can be grouped into the $f(R)$-family theories \cite{suvorov/2019}-\cite{santos/2010} and its extensions, namely $f(R,L)$ \cite{harko/2010,azevedo/2016} and $f(R,T)$ theories of gravity \cite{harko/2011}-\cite{shabani/2013}, with $R$, $L$ and $T$ being the Ricci scalar, matter lagrangian density and trace of the energy-momentum tensor, respectively. In these theories, $R$ in the Einstein-Hilbert action,
	
	\begin{equation}\label{i3}
		S=\frac{c^4}{16\pi G}\int d^4x\sqrt{-g}R
	\end{equation}
	is substituted by a generic function $f$ (with the argument being dependent on the theory one is working with). In (\ref{i3}), $g$ is the metric determinant.
	
	Another examples are the Gauss-Bonnet-family theories \cite{giribet/2006}-\cite{de_felice/2009}, in which the Gauss-Bonnet scalar, $\mathcal{G}$, or even a function of it, $f(\mathcal{G})$, is inserted in (\ref{i3}).  
	
	Note that there is also the possibility of working with a function of both $R$ and $\mathcal{G}$ in the Einstein-Hilbert action, what contemplates the $f(R,\mathcal{G})$ theories \cite{de_felice/2010}-\cite{santos_da_costa/2018}.
	
	There are also theories that investigate the role of the torsion $\mathcal{T}$ in gravity or, once again, a general function of it, namely $f(\mathcal{T})$. Those are the teleparallel gravity theories \cite{cai/2016}-\cite{de_andrade/2000}, which also have their extensions \cite{harko/2014}-\cite{yousaf/2018}.
	
	There are many other forms of constructing cosmic acceleration models, through quintessence \cite{mak/2002}-\cite{linder/2008}, Chaplygin gas EoS \cite{bento/2003}-\cite{chakraborty/2007} and even extradimensional models \cite{moraes/2016b}.
	
	With a host of alternatives able to describe the cosmic acceleration in agreement with cosmological observations in the present literature (check also \cite{trodden/2007}-\cite{yang/2019}), unless higher precision observational data regarding the universe expansion rate is available, we cannot distinguish one alternative from the other in the search for the ultimate cosmological model.
	
	A more conservative approach, relying on as few model-dependent quantities as possible, could be the remedy to this current and persistent issue. Rather than finding and solving Friedmann-like equations for a plethora of cosmological models, one could invoke cosmography \cite{weinberg/1972}-\cite{visser/2005}.
	
	A large amount of modern cosmology is surprisingly pure kinematics and is completely independent of the underlying dynamics governing the evolution of the universe, described by the Friedmann equations. Recalling that the scale factor in Friedmann-Lem\^aitre-Robertson-Walker metric dictates how distances evolve in the universe (check next section), cosmography does not predict anything about the scale factor, but, to some extent, it can infer its history from observational data. 
	
	Cosmography is only related to the derivatives of the scale factor, that generate the cosmographic parameters. Those make it possible to fit the distance-redshift relation data without any assumption on the underlying cosmological model, that is, cosmography does not rely on the solution of cosmic equations. Once the cosmographic parameters are determined, one could use them to put constraints on a given cosmological model, by writing the model characterizing quantities as functions of such parameters.
	
	Cosmography has already been developed to Extended Gravity Theories, particularly to the $f(R)$ gravity \cite{pannia/2013}-\cite{capozziello/2008b} (the last of these references was actually applied to an $f(R)$-brane cosmology) and $f(\mathcal{T})$ gravity \cite{capozziello/2011b,piedipalumbo/2015}. For a review of these approaches, check \cite{capozziello/2019}.  
	
	As far as the present authors know, despite the large number of applications that the $f(R,T)$ theory of gravity currently presents (besides \cite{harko/2011}-\cite{shabani/2013}, one can also check, for instance, \cite{barrientos/2018}), there is no $f(R,T)$ cosmography. In this manner, the main goal of the present article is to obtain the cosmography of the $f(R,T)$ gravity theory.
	
	\section{Cosmography and cosmographic apparatus}\label{sec:cca}
	
	Let us conveniently start by writing the well-known Friedmann-Lem\^aitre-Robertson-Walker  metric,
	
	\begin{equation}\label{cca1}
		ds^2=-dt^2+a(t)^2\left [ \frac{dr^2}{1-kr^2}+r^2\left ( d\theta^2+sin^2\theta d\phi^2 \right ) \right ],
	\end{equation}
	that represents a homogeneous and isotropic expanding universe with spatial curvature $k$, whose expansion is driven by the aforementioned scale factor $a(t)$. In \cite{vincenzo/2010}, it was shown that ranging the curvature parameter $k$ in the interval $[-1,+1]$ has negligible effects on the estimation of the cosmographic parameters. Anyhow, observational outcomes suggest a null spatial curvature \cite{cai/2016b}, so that, in the following, we will assume a flat universe and use $k=0$. %Thus we have to (\ref{cca1}) that
	
	%\begin{equation} \label{cca2}
	%ds^2=-dt^2+a(t)^2\left [ dr^2+r^2\left ( d\theta^2+sin^2\theta d\phi^2 \right ) \right ].
	%\end{equation}
	
	%In observational cosmology we do not have direct access to the complete history of $a(t)$. We only have access to the current value of the scale factor and its derivatives, which are encoded in the Hubble parameter, deceleration parameter etc. 
	
	Expanding the scale factor in Taylor series around $t_0$ leads to a distance-redshift relation which only relies on the assumption of metric (\ref{cca1}), being fully model-independent:
	
	%\begin{equation} \label{cca3}
	%a\left ( t \right )=\sum_{n=0}^{\infty }\frac{1}{n!}\frac{d^na\left ( t \right )}{dt^n}\mid_{t_{0}}\left ( t-t_{0} \right )^n,
	%\end{equation}
	%in fifth order we have to
	
	\begin{equation} \label{cca4}
		a(t)=a_0+\dot{a}(t_0)(t-t_0)+\frac{1}{2!}\ddot{a}(t_0)(t-t_0)^2+\frac{1}{3!}\dddot{a}(t_0)(t-t_0)^3+\frac{1}{4!}\ddddot{a}(t_0)(t-t_0)^4.
	\end{equation}	
	
	Throughout this article the subscript $0$ indicates the present value of the quantity, a dot denotes the derivative with respect to the cosmic time and terms of order $\geq 5$ were neglected in (\ref{cca4}).
	
	Now, let us define a set of functions in terms of the derivatives of the scale factor \cite{weinberg/1972}-\cite{visser/2005}:
	
	\begin{equation} \label{cca5}
		H(t)=\frac{\dot{a}(t)}{a(t)},
	\end{equation}
	
	\begin{equation} \label{cca6}
		q(t)=-\frac{\ddot{a}(t)}{a(t)}H^{-2},
	\end{equation}
	
	\begin{equation} \label{cca7}
		j(t)=\frac{\dddot{a}(t)}{a(t)}H^{-3},
	\end{equation}
	
	\begin{equation} \label{cca8}
		s(t)=\frac{\ddddot{a}(t)}{a(t)}H^{-4},
	\end{equation}
	which are usually referred to as Hubble, deceleration, jerk, and snap parameters, respectively. 
	%\begin{equation} \label{cca9}
	%l(t)=\frac{1}{a(t)}\frac{d^5a(t)}{dt^5}H^{-5},
	%\end{equation}
	%(see \cite{maciej/2008} for a historical account of these names). 
	
	Using these definitions it is easy to rewrite (\ref{cca4}) as 
	
	\begin{equation} \label{cca10}
		a(t)=a_0\left[1+H_0(t-t_0)-\frac{1}{2!}q_0H_0^2(t-t_0)^2+\frac{1}{3!}j_0H_0^3(t-t_0)^3+\frac{1}{4!}s_0H_0^4(t-t_0)^4\right].
	\end{equation}	
	
	Using the definition of the Hubble function (\ref{cca5}), as well as the definitions of the others cosmographic parameters (\ref{cca6})-(\ref{cca8}), we can demonstrate the following relationships among the derivatives of the Hubble function and the cosmographic parameters
	
	\begin{equation} \label{cca11}
		\dot{H}=-H^2(1+q),
	\end{equation}
	
	\begin{equation} \label{cca12}
		\ddot{H}=H^3\left ( j+3q+2 \right ), 
	\end{equation}
	
	\begin{equation} \label{cca13}
		\dddot{H}=H^4\left [ s-4j-3q(q+4) -6\right ],
	\end{equation}
	
	%\begin{equation} \label{cca14}
	%\ddddot{H}=H^5\left [ l-5s+10\left ( q+2 \right ) +30\left ( q+2 %\right )q+24\right ], 
	%\end{equation}
	
	It is also possible to express the Ricci curvature scalar $R$ in terms of the Hubble function as %\cite{capozziello/2005} 
	
	\begin{equation} \label{cca15}
		R=-6\left ( \dot{H}+2H^2 \right ).
	\end{equation}
	Differentiating this with respect to $t$, we obtain
	
	\begin{equation} \label{cca16}
		\dot{R}=-6\left ( \ddot{H}+4H\dot{H} \right ),
	\end{equation}
	
	\begin{equation} \label{cca17}
		\ddot{R}=-6\left ( \dddot{H}+4H\ddot{H}+ 4\dot{H}^2\right ),
	\end{equation}
	
	%\begin{equation} \label{cca18}
	%\dddot{R}=-6\left ( d^4H/dt^4+4\dddot{H}H+12\dot{H}\ddot{H}\right ).
	%\end{equation}
	
	Evaluating these expressions at present time $t_0$ and using Equations (\ref{cca11})-(\ref{cca13}), one finally gets 
	
	\begin{equation} \label{cca19}
		R_0=-6H_0^2\left ( 1-q_0 \right ),
	\end{equation}
	
	\begin{equation} \label{cca20}
		\dot{R_0}=-6H_0^3\left ( j_0-q_0-2 \right ),
	\end{equation}
	
	\begin{equation} \label{cca21}
		\ddot{R_0}=-6H_0^4\left (s_0+ q_0^2+8q_0+6 \right ),
	\end{equation}
	
	%\begin{equation} \label{cca22}
	%\dddot{R_0}=-6H_0^5\left %[l_0-s_0+2(q_0+4)j_0-6(3q_0+8)q_0-24\right].
	%\end{equation}
	The relations (\ref{cca11})-(\ref{cca13}) and (\ref{cca19})-(\ref{cca21}) will turn out to be useful in the following.
	
	\section{Fundaments of the $f(R,T)$ gravity theory}\label{sec:frt}
	
	The $f(R,T)$ gravity theory starts from the total action \cite{harko/2011}
	
	\begin{equation}\label{frt1}
		S=\int d^4x\sqrt{-g}\left[\frac{1}{16\pi G}f(R,T)+L\right],
	\end{equation}
	assuming $c=1$. The motivation for inserting terms depending on $T$ in the gravitational action above will be carefully discussed below.
	
	The explicit dependence of the theory on $T$ may be due to: ${\bf i)}$ {\it quantum effects}. The quantum effects contained in $T$ may be related to the mechanism of particle production since those terms prevent the energy-momentum tensor of the theory to be conserved, as it is going to be shown below; ${\bf ii)}$ {\it imperfect fluid}. In this sense, imperfection can be led by viscosity or anisotropy. Recall that anisotropy arises naturally in matter field at high densities \cite{ruderman/1972,canuto/1974}; ${\bf iii)}$ {\it extra fluid}. The extra terms proportional to $T$ may represent an extra fluid permeating the studied system together with the usual fluid \cite{moraes/2018}. Everything happens as if the system was permeated by one fluid whose density (pressure) is given by the sum of the extra and usual densities (pressures); ${\bf iv)}$ {\it varying cosmological ``constant''}. It has been shown that a cosmological model derived from the $f(R,T)=R+2\lambda T$ gravity, with constant $\lambda$, in a pressureless matter dominated universe is equivalent to a model with effective cosmological constant $\Lambda(t)\sim H^2$ \cite{harko/2011}. Note that some running vacuum models, such as the one presented in \cite{geng/2017}, also have $\Lambda(t)\propto H^2$, but such a dependency is assumed from phenomenological reasons.
	
	By varying action (\ref{frt1}) with respect to the metric yields the following field equations:
	
	\begin{equation}\label{frt2}
		f_RR_{\mu\nu}-\frac{1}{2}fg_{\mu\nu}+(g_{\mu\nu}\Box-\nabla_\mu\nabla_\nu)f_R=8\pi GT_{\mu\nu}+f_T(T_{\mu\nu}-Lg_{\mu\nu}),
	\end{equation}
	with $f=f(R, T)$, $f_R=f_R(R,T)\equiv\partial f(R,T)/\partial R$, $R_{\mu\nu}$ is the Ricci tensor and $f_T=f_T(R,T)\equiv\partial f(R,T)/\partial T$.
	
	The covariant derivative of Eq.(\ref{frt2}) can be written as
	
	\begin{equation}\label{frt3}
		\nabla^\mu T_{\mu\nu}=\frac{f_T}{8\pi+f_T}\left[(Lg_{\mu\nu}-T_{\mu\nu})\nabla^\mu\ln f_T(R,T)+\nabla^\mu\left(L-\frac{1}{2}T\right)g_{\mu\nu}\right].
	\end{equation}

	Eq.(\ref{frt3}) shows that a general $f(R,T)$ function does not conserve the energy-momentum tensor. This can be interpreted as due to the creation (or destruction) of particles throughout the universe evolution \cite{singh/2016}-\cite{asadiyan/2019}. Nevertheless, in \cite{dos_santos/2019}-\cite{chakraborty/2013} e.g. some conservative models in $f(R,T)$ gravity were obtained.
	
	\section{Cosmography of the $f(R,T)$ gravity theory}\label{sec:cfrt}
	
	Applying the Friedmann-Lem\^aitre-Robertson-Walker metric (\ref{cca1}) to the field equations (\ref{frt2}) with $L=p$ for a perfect fluid such that $T^{\mu \nu}=diag(-\rho , p, p, p)$ gives \cite{shabani/2013} 
	
	\begin{equation} \label{cfrt1}
		3H^{2}f_R+\frac{1}{2}\left ( f-f_RR \right)+3\dot{f}_RH=\left ( 8\pi G+f_T \right )\rho +f_Tp,
	\end{equation}
	as the modified Friedmann equation and 
	
	\begin{equation} \label{cfrt2}
		2f_R\dot{H}+\ddot{f}_R-\dot{f}_RH=-\left ( 8\pi G+ f_T \right )\left( \rho +p\right ), 
	\end{equation}
	as the modified Raychaudhuri equation. 
	
	Now considering the expressions (\ref{cfrt1}) and (\ref{cfrt2}) for the case of a flat universe filled with dust, that is $p=0$, we have, respectively,
	
	\begin{equation} \label{cfrt3}
		3H^{2}f_R+\frac{1}{2}\left ( f-f_RR \right)+3\dot{f}_RH=\left ( 8\pi G+f_T \right )\rho, 
	\end{equation}
	
	\begin{equation} \label{cfrt4}
		2f_R\dot{H}+\ddot{f}_R-\dot{f}_RH=-\left ( 8\pi G+ f_T \right )\rho.
	\end{equation}
	Expressions (\ref{cfrt3}) and (\ref{cfrt4}) form a system with two equations and five unknowns, namely: $f$, $f_T$, $f_R$, $\dot{f}_R$ and $\ddot{f}_R$. In order to make this system closed, we need to make some assumptions, which may be different for different classes of models $f(R, T)$. So from now on, we will work with the class $f(R, T)= f(R) + f(T)$, where $f(R)$ and $f(T)$ are two arbitrary functions of $R$ and $T$ respectively.
	
	In the previous section we saw that in general, $f(R,T)$ gravity models do not conserve the energy-momentum tensor, although some conservative models have already been proposed within this approach. In \cite{velten/2017} it was shown that cosmological models based on conservative $f(R,T)$ gravity agrees with observational data. In what follows, we will assume that our $f(R,T)$ gravity model conserves the energy-momentum tensor, so that we can write \cite{capozziello/2005}
	
	\begin{equation} \label{cfrt5}
		\rho=3H_0\frac{\Omega_m}{a^{3}},
	\end{equation}
	where $\Omega_m$ is the dimensionless matter density parameter. Furthermore, from now on we are assuming that $8\pi G=1$. 
	
	Equations (\ref{cfrt3}) and (\ref{cfrt4}) now read respectively as
	
	\begin{equation} \label{cfrt6}
		3H^{2}f_R+\frac{1}{2}\left ( f-f_RR \right)+3\dot{f}_RH=\left ( 1+f_T \right )3H_0\frac{\Omega_m}{a^{3}}, 
	\end{equation}
	
	\begin{equation} \label{cfrt7}
		2f_R\dot{H}+\ddot{f}_R-\dot{f}_RH=-\left ( 1+ f_T \right )3H_0\frac{\Omega_m}{a^{3}}.
	\end{equation}
	
	In order to be able to close the system of equations, we differentiate (\ref{cfrt7}) with respect to time to get
	
	\begin{equation} \label{cfrt8}
		\ddot{H}= \frac{\left ( 9H_{0}^{2}\frac{\Omega_mH}{a^{3}}-\dddot{f}_R+\ddot{f}_RH+\dot{f}_R\dot{H} \right)}{2f_R}
		+\frac{\left ( 3H_{0}^{2}\frac{\Omega_m}{a^{3}}+\ddot{f}_R-\dot{f}_RH \right )\dot{f}_R}{2f_R^2}.
	\end{equation}
	
	Now from the field equations of this class of $f(R,T)$ models the gravitational constant has an effective value given by \cite{harko/2011}
	
	\begin{equation} \label{cfrt9}
		G_{eff}=\frac{1}{{f}'\left ( R \right )}\left [ 1+\frac{{f}'\left ( T \right )}{8 \pi} \right ],
	\end{equation}
	where the line denotes differentiation with respect to the argument. In a good approximation we must have in $t=t_0$ that $G_{eff}\longrightarrow G$. From Equation (\ref{cfrt9}) this implies in
	
	\numparts
	\begin{eqnarray} \label{cfrt10}
		{f}'\left ( T_0 \right )=0,\\ 
		{f}'\left ( R_0 \right )=1.
	\end{eqnarray}
	\endnumparts
	This in turn, due to the particular form of $f(R,T)$ assumed here, implies respectively that
	
	\numparts
	\begin{eqnarray} \label{cfrt11}
		f^{(0)}_T=0,  \\ 
		f^{(0)}_R=1,
	\end{eqnarray}
	\endnumparts
	where the superscript $0$ also means that the derivative is being taken in $t_0$. 
	
	By using this result together with the fact that in $t_0$ the scale factor is established as being equal to the unit, $a_0=1$, we have for Equations (\ref{cfrt6}), (\ref{cfrt7}) and (\ref{cfrt8}) respectively that
	
	\begin{equation} \label{cfrt12}
		3H_0^2+\frac{1}{2}\left ( f_0-R_0 \right)+3\dot{f}_R^{(0)}H_0=3H_0\Omega_m,
	\end{equation}
	
	\begin{equation} \label{cfrt13}
		2\dot{H}_0+\ddot{f}_R^{(0)}-\dot{f}_R^{(0)}H_0=-3H_0\Omega_m,
	\end{equation}
	
	\begin{eqnarray} \label{cfrt14}
		\ddot{H}_0= \frac{\left ( 9H_{0}^{3}\Omega_m-\dddot{f}_R^{(0)}+\ddot{f}_R^{(0)}H_0+\dot{f}_R^{(0)}\dot{H}_0 \right)}{2} \\ \nonumber
		+\frac{\left ( 3H_{0}^{2}\Omega_m+\ddot{f}_R^{(0)}-\dot{f}_R^{(0)}H_0 \right )\dot{f}_R^{(0)}}{2}.
	\end{eqnarray}
	
	Using some mathematical artifices, we can write the time derivatives as derivatives with respect to the argument, so that Equations (\ref{cfrt12}), (\ref{cfrt13}) and (\ref{cfrt14}) become respectively
	
	\begin{equation} \label{cfrt15}
		3H_0^2+\frac{1}{2}\left ( f_0-R_0 \right)+3\dot{R}_0f_{RR}^{(0)}H_0=3H_0\Omega_m, 
	\end{equation}
	
	\begin{equation} \label{cfrt16}
		2\dot{H}_0+\ddot{R}_0f_{RRR}^{(0)}-\dot{R}_0f_{RR}^{(0)}H_0=-3H_0\Omega_m,
	\end{equation}
	
	\begin{eqnarray}\label{cfrtx}
		\ddot{H}_0= \frac{\left( 9H_{0}^{3}\Omega_m-\dddot{R}_0f_{RRRR}^{(0)}+\ddot{R}_0f_{RRR}^{(0)}H_0+\dot{R}_0f_{RR}^{(0)}\dot{H}_0 \right)}{2} \\ \nonumber
		+\frac{\left ( 3H_{0}^{2}\Omega_m+\ddot{R}_0f_{RRR}^{(0)}-\dot{R}_0f_{RR}^{(0)}H_0 \right )\dot{R}_0f_{RR}^{(0)}}{2}. 
	\end{eqnarray}
	
	Finally, let us assume that the function $f(R)$ may be well approximated by its third order Taylor expansion in $R-R_0$, in such a way that $d^nf(R)/dR^n=0$ for $n\geq 4$ \cite{capozziello/2008}. In such an approximation, $f_{RRRR}^{(0)}=0$ so that we have for (\ref{cfrtx}) that
	
	\begin{eqnarray} \label{cfrt17}
		\ddot{H}_0= \frac{ 9H_{0}^{3}\Omega_m+\ddot{R}_0f_{RRR}^{(0)}H_0+\dot{R}_0f_{RR}^{(0)}\dot{H}_0}{2} \\ \nonumber
		+\frac{\left ( 3H_{0}^{2}\Omega_m+\ddot{R}_0f_{RRR}^{(0)}-\dot{R}_0f_{RR}^{(0)}H_0 \right )\dot{R}_0f_{RR}^{(0)}}{2}.
	\end{eqnarray}
	
	Now we have that Equations (\ref{cfrt15}), (\ref{cfrt16}) and (\ref{cfrt17}) form a closed system with three equations and three unknowns, namely: $f_0$, $f_{RR}^{(0)}$ and $f_{RRR}^{(0)}$. This system of equations can be solved, and then using (\ref{cca11})-(\ref{cca12}) and (\ref{cca19})-(\ref{cca21}) we can find expressions for $f_0$, $f_{RR}^{(0)}$ and $f_{RRR}^{(0)}$ in terms of the cosmographic parameters $(q_0, j_0, s_0)$. By doing so we obtain
	
	%Solving the system formed by (\ref{cfrt15}), (\ref{cfrt16}) and (\ref{cfrt17}) for the unknowns, and using the expressions (\ref{cca11})-(\ref{cca13}) and (\ref{cca19})-(\ref{cca21}), one ends up with the equations that relate $f_0$, $f_{RR}^{(0)}$ and $f_{RRR}^{(0)}$ to the cosmographic set $(q_0, j_0, s_0)$ respectively as 
	
	\begin{equation} \label{cfrt18}
		\frac{f_0}{6H_0}=\frac{\mathcal{P}_0(H_0)\Omega_M^2+\mathcal{Q}_0(q_0, H_0)\Omega_M-\mathcal{S}_0(q_0, j_0, H_0)}{\mathcal{R}(\Omega_M, q_0, H_0)},
	\end{equation}
	
	\begin{equation} \label{cfrt19}
		\frac{f_{RR}^{(0)}}{\left ( H_0^2 \right )^{-1}}=\frac{\mathcal{P}_2(H_0)\Omega_M+\mathcal{Q}_2(q_0, j_0, H_0)}{\left ( 6j_0-6q_0-12 \right )\mathcal{R}(\Omega_M, q_0, H_0)},
	\end{equation}
	
	\begin{equation} \label{cfrt20}
		\frac{f_{RRR}^{(0)}}{\left ( H_0^2 \right )^{-2}}=\frac{\mathcal{P}_3(H_0)\Omega_M^2+\mathcal{Q}_3(q_0, H_0)\Omega_M-\mathcal{S}_3(q_0, j_0, H_0)}{\left ( 6s_0+6q_0^2+48q_0+36 \right )\mathcal{R}(\Omega_M, q_0, H_0)},
	\end{equation}
	where we have defined the following quantities
	
	\begin{equation} \label{cfrt21}
		\mathcal{P}_0(H_0)=3\left (H_0 -1\right ),
	\end{equation}
	
	\begin{equation} \label{cfrt22}
		\mathcal{Q}_0(q_0, H_0)=\left[\left ( 5-2q_0 \right )+3\left ( 1+q_0 \right )H_0 \right]H_0,
	\end{equation}
	
	\begin{equation} \label{cfrt23}
		\mathcal{S}_0(q_0, j_0, H_0)=\left[2j_0-(q_0-4)q_0+6 \right ]H_0^2,
	\end{equation}
	
	\begin{equation} \label{cfrt24}
		\mathcal{P}_2(H_0)=3\left(1-3H_0 \right),
	\end{equation}
	
	\begin{equation} \label{cfrt25}
		\mathcal{Q}_2(q_0, j_0, H_0)=2\left(j_0+2q_0+1 \right)H_0,
	\end{equation}
	
	\begin{equation} \label{cfrt26}
		\mathcal{P}_3(H_0)=9\left(H_0-1 \right )H_0,
	\end{equation}
	
	\begin{equation} \label{cfrt27}
		\mathcal{Q}_3(q_0, H_0)=3\left[3(1+q_0)+(1-2q_0)H_0 \right ]H_0^2,
	\end{equation}
	
	\begin{equation} \label{cfrt28}
		\mathcal{S}_3(q_0, j_0, H_0)=2\left[j_0+\left ( 5+q_0 \right )q_0+3\right]H_0^3,
	\end{equation}
	
	\begin{equation} \label{cfrt29}
		\mathcal{R}(\Omega_M, q_0, H_0)=\left[-3\Omega_M+\left (2+q_0+3\Omega_M\right)H_0\right ].
	\end{equation}
	The Equations (\ref{cfrt18})-(\ref{cfrt29}) make it possible to estimate the present-day values of $f(R, T)$ and its three first derivatives as  functions of the Hubble constant $H_0$ and the cosmographic set $(q_0, j_0, s_0)$ provided a value for the matter density parameter $\Omega_m$.
	
	\section{Results}\label{sec:rd} 
	
	\subsection{$f(R, T)$ theories and CPL dark energy models} \label{subsrd1}
	
	It is well known in the literature that $f(R, T)$ gravity models can evade the cosmological constant problem since such models allow to obtain the current accelerated expansion of the universe through extra terms that appear in the field equations \cite{barrientos/2014, kumar/2015,singh/2016,moraes/2016c,moraes/2016d,moraes/2017,moraes/2018}. On the other hand, we have that cosmographic parameters can also be expressed in terms of dark energy density and dark energy EoS parameters, so that we can solve the cosmographic equations in expressions (\ref{cfrt18})-(\ref{cfrt20}) giving the same set $(q_0, j_0, s_0)$ as the dark energy model considered. 
	
	To this aim, following the prescription of the so-called Dark Energy Task Force \cite{taskforce}, we will use the CPL (Chevallier-Polarski-Linder) parametrization for the EoS, which is a parametrized expression developed in \cite{chevallier, linder/2003}, namely
	
	\begin{equation} \label{rd1}
		w=w_0+w_a(1-a)=w_0+w_az(1+z)^{-1},  
	\end{equation}
	where $w_0$ denotes the present value of the EoS, while $w_a$ is a constant and $z$ is the redshift. Because of its bounded behavior at high redshift and high accuracy in reconstructing many scalar field EoS, the CPL parametrization has become one of the most popular methods to study dark energy. In addition, this parametrized expression is convenient in order to reduce the dependence of the results on any underlying theoretical scenario. Thus in a flat universe filled by dust matter and dark energy, we can write the dimensioneless Hubble parameter $E(z)\equiv H(z)/H_0 $ as \cite{capozziello/2008b}
	
	\begin{equation} \label{rd2}
		E^2(z)=\Omega_M(1+z)^3+(1-\Omega_M)(1+z)^{3(1+w_0+w_a)}e^{-(3w_az/1+z)}.
	\end{equation}
	
	Here, instead of integrating the expression (\ref{rd2}) for $H(z)$ to get $a(t)$, in order to determine the cosmographic parameters for such a model, we can use the relation 
	
	\begin{equation} \label{rd3}
		\frac{d}{dt}=-(1+z)H(z)\frac{d}{dz}  
	\end{equation}
	to evaluate $(\dot{H}, \ddot{H}, \dddot{H})$ and then solve the expressions (\ref{cca11})-(\ref{cca13}) with respect to the parameters of interest, evaluated in $z=0$. Some algebra gives \cite{capozziello/2011c}
	
	\begin{eqnarray} \label{rd4}
		q_0=\frac{1}{2}+\frac{3}{2}(1-\Omega_M)w_0,
	\end{eqnarray}
	
	\begin{eqnarray} \label{rd5}
		j_0=1+\frac{3}{2}(1-\Omega_M)\left [ 3w_0(1+w_0)+w_a \right ],
	\end{eqnarray}
	
	\begin{eqnarray} \label{rd6}
		s_0=-\frac{7}{2}-\frac{33}{4}(1-\Omega_M)w_a-\frac{9}{4}(1-\Omega_M)\\ \nonumber
		\times \left [ 9+(7-\Omega_M)w_a \right ]w_0-\frac{9}{4}(1-\Omega_M)
		\\ \nonumber
		\times (16-3\Omega_M)w_0^2-\frac{27}{4}(1-\Omega_M)(3-\Omega_M)w_0^3.
	\end{eqnarray}
	
	%\begin{eqnarray} \label{rd7}
	%l_0=\frac{35}{2}+\frac{(1-\Omega_M)}{4}\left [ 213+(7-\Omega_M) w_a\right ]w_a \nonumber \\
	%+\frac{(1-\Omega_M)}{4}\left [489+9(82-21\Omega_M)w_a \right ]w_0 \nonumber \\
	%+\frac{9}{2}(1-\Omega_M)\left [ 67-21\Omega_M+\frac{3}{2} \left ( 23-11\Omega_M \right )w_a\right ]w_0^2 \nonumber \\
	%+\frac{27}{4}(1-\Omega_M)(47-24\Omega_M)w_0^3 \nonumber \\
	%+\frac{81}{2}(1-\Omega_M)(3-2\Omega_M)w_0^4.
	%\end{eqnarray}
	
	Once we have the expressions (\ref{rd4})-(\ref{rd6}) for the cosmographic parameters in terms of the CPL model parameters, we can use the values of $(w_0, w_a)$ for some particular dark energy models and thus rewrite the cosmographic equations for these models.
	
	\subsubsection{$\Lambda$CDM model}
	
	As a first case, we will consider the $\Lambda$CDM model (also known as standard model, with  CDM denoting {\it cold dark matter}), for which $(w_0, w_a)=(-1, 0)$ \cite{bamba/2012}, so that the cosmographic parameters given by (\ref{rd4})-(\ref{rd6}) become respectively
	
	\numparts
	\begin{eqnarray} \label{rd8}
		q_0=\frac{1}{2}-\frac{3}{2}(1-\Omega_M),\\
		j_0=1,\\
		s_0=1-\frac{9}{2}\Omega_M.
	\end{eqnarray}
	\endnumparts
	Therefore, inserting the values of $(q_0, j_0, s_0)$ given in (58), we have for the equations (\ref{cfrt18})-(\ref{cfrt20}) respectively that
	
	\begin{equation} \label{rd9}
		f_0=\frac{3H_0\left [ -12H_0-8H_0\Omega_M+3(-4+3H_0+6H_0^2) \right ]\Omega_M^2}{\mathcal{M}},
	\end{equation}
	
	\begin{equation} \label{rd10}
		f_{RR}^{(0)}=\frac{2\left ( H_0-1 \right )}{3H_0^2\mathcal{M}},
	\end{equation}
	
	\begin{equation} \label{rd11}
		f_{RRR}^{(0)}=\frac{2(H_0-1)(3H_0-2)\Omega_M^2}{3H_0^3(2+\Omega_M)\mathcal{M}}.
	\end{equation}
	In (\ref{rd9})-(\ref{rd11}), in order to be able to write the expressions in a shorter way, we have defined the following quantity
	
	\begin{equation} \label{rd12}
		\mathcal{M}=(2+9\Omega_M)H_0-6\Omega_M.
	\end{equation}
	Therefore, by assigning values to the Hubble constant $H_0$ and to the matter density parameter $\Omega_m$, we can obtain numerical expressions for $f_0$, $f_{RR}^{(0)}$ and $f_{RRR}^{(0)}$ within the $\Lambda$CDM scenario.
	
	\subsubsection{\textit{Quiessence} models}
	
	As a second case of dark energy model, we will now consider the quiessence models, which are models with rolling fields. They are direct generalizations of the cosmological constant, that is,  when the fields are not rolling, their potential energy behaves as a cosmological constant, with $w=constant$ \cite{bamba/2012}. Naturally for $w_0 = -1$ we recover the previous $\Lambda$CDM case. For these models we have for (\ref{rd4})-(\ref{rd6}) respectively that
	
	\numparts
	\begin{eqnarray} \label{rd13}
		q_0=\frac{1}{2}+\frac{3}{2}(1-\Omega_M)w_0,\\  j_0=1+\frac{3}{2}(1-\Omega_M)(3w_0+3w_0^2), \\ 
		s_0=-\frac{7}{2}-\frac{81}{4}(1-\Omega_M)w_0-\frac{9}{4}(1-\Omega_M)(16-3\Omega_M)w_0^2 \\ \nonumber
		-\frac{27}{4}(1-\Omega_M)(3-\Omega_M)w_0^3.
	\end{eqnarray}
	\endnumparts
	Substituting these values of $(q_0, j_0, s_0)$ in (\ref{cfrt18})-(\ref{cfrt20}), we have that the cosmographic equations of the $f(R, T)$ for the quiessence models are respectively
	
	\begin{eqnarray} \label{rd14}
		f_0=\frac{3H_0\left \{3H_0\left [ 13+9w_0(2+w_0) \right ]-2H_0\left [ 8+9H_0(1+w_0)+3w_0(7+3w_0) \right ]\Omega_M\right \}}{\mathcal{N}} \nonumber \\
		+\frac{3H_0\left \{3\left [4+H_0\left ( -4+(-4+6H_0-3w_0)w_0 \right)\right ]\Omega_M^2\right \}}{\mathcal{N}},
	\end{eqnarray}
	\begin{equation} \label{rd15}
		f_{RR}^{(0)}=\frac{2\left \{ H_0\left [ 2-w_0(5+3w_0)\mathcal{M}_1-3\Omega_M \right ]+\Omega_M \right \}}{3H_0^2\left [ 1+w_0(2+3w_0)\mathcal{M}_1 \right ]\mathcal{N}},
	\end{equation}
	
	\begin{equation} \label{rd16}
		f_{RRR}^{(0)}=\frac{-6H_0^2(1+w_0)^2+2H_0(1+w_0)(3+4H_0w_0)\Omega_M}{\left \{ -3+w_0\mathcal{M}_1\left [ -3+w_0(-15+3w_0\mathcal{M}_3+2\Omega_M) \right ] \right \}\mathcal{N}}.
	\end{equation}
	In order to write the above equations in a shorter way, the following quantities have been defined
	
	\begin{eqnarray} \label{rd17}
		\mathcal{N}=\left [ 3w_0(\Omega_M-1)-6\Omega_M-5 \right ]H_0+6\Omega_M,\\ 
		\mathcal{M}_1=\Omega_M-1,\\ 
		\mathcal{M}_3=\Omega_M-3.
	\end{eqnarray}
	Therefore, by assigning values for $H_0$, $\Omega_m$ and $w_0$, we can obtain numerical expressions for the $f(R, T)$ cosmography in this scenario of quiessence.
	
	\subsubsection{Evolving dark energy models}
	
	Finally, we will consider here a general model of dark energy with evolving EoS, that is, $w_a \neq 0$. Furthermore, in agreement with some of the most recent analyses \cite{bamba/2012}, we only concentrate on evolving dark energy models with $w_0=-1$. The cosmographic parameters for such models are
	
	\numparts
	\begin{eqnarray} \label{rd18}
		q_0=\frac{1}{2}-\frac{3}{2}(1-\Omega_M),\\  j_0=1+\frac{3}{2}(1-\Omega_M)w_a, \\
		s_0=-\frac{7}{2}-\frac{33}{4}(1-\Omega_M)w_a+\frac{9}{4}(1-\Omega_M)\left [ 9+(7-\Omega_M)w_a \right] \\ 
		-\frac{9}{4}(1-\Omega_M)(16-3\Omega_M)+\frac{27}{4}(1-\Omega_M)(3-\Omega_M). \nonumber
	\end{eqnarray}
	\endnumparts
	Inserting the values of $(q_0, j_0, s_0)$ given in (70) into (\ref{cfrt18})-(\ref{cfrt20}), we have the $f(R, T)$ cosmography for these models:
	
	\begin{equation} \label{rd19}
		f_0=\frac{3H_0\left [ -12H_0(1+w_a)+4H_0(-2+3w_a)\Omega_M+3(-4+3H_0+6H_0^2)\Omega_M^2\right ]}{\mathcal{M}},
	\end{equation}
	
	\begin{equation} \label{rd20}
		f_{RR}^{(0)}=\frac{2\left \{ -\Omega_M+H_0\left [ w_a\mathcal{M}_1+\Omega_M \right ] \right \}}{3\left ( w_a\mathcal{M}_1+\Omega_M \right )\mathcal{M}},
	\end{equation}
	
	\begin{equation} \label{rd21}
		f_{RRR}^{(0)}=\frac{-4H_0^3w_a(1-\Omega_M)-6(H_0-1)H_0(3H_0-2)\Omega_M^2}{3\left [ 3\Omega_M(2+\Omega_M)+w_a\mathcal{M}_1(-10+3\Omega_M)\right ]\mathcal{M}}.
	\end{equation}
	Thus, by assigning values for $H_0$, $\Omega_m$ and $w_a$, we can obtain numerical expressions for $f_0$, $f_{RR}^{(0)}$ and $f_{RRR}^{(0)}$ for this case of dark energy model with EoS in evolution.
	
	This procedure to estimate $f(R, T)$ and its derivatives resorting to a model independent parametrization of the dark energy EoS allows us to write the present-day values of these quantities as a function of $(\Omega_m, w_0, w_a)$, so that, using values for these parameters, numerical expressions could be obtained for $f_0$, $f_{RR}^{(0)}$ and $f_{RRR}^{(0)}$. An analysis of the space of these parameters for which it would be possible to obtain reliable models can be done numerically, but this is outside the scope of this article..  It is worth noting that the $f(R, T)$ model thus obtained is not dynamically equivalent to the starting CPL one. Indeed, the two models have the same cosmographic parameters only today \cite{capozziello/2008b, bamba/2012}. In the following, we will show how cosmographic equations (\ref{cfrt18})-(\ref{cfrt20}) can be used to constrain viable $f(R, T)$ models from the observational values of $(q_0, j_0, s_0)$.
	
	\subsection{Restricting $f(R,T)$ gravity models using cosmography} \label{subsrd2}
	
	In order to use cosmographic equations to restrict some particular models of $f(R, T)$ theories, we must proceed, basically, in three steps. Those are:
	
	$1$ - we write some model of the form $f(R, T) = f(R) + f(T)$ in terms of general parameters;
	
	$2$ - we obtain the derivatives of this particular $f(R, T)$ function in terms of the general parameters mentioned above;
	
	$3$ - as the values of the $f(R, T)$ function and its derivatives depend on the cosmographic parameters $(q_0, j_0, s_0)$, we have that the general parameters that characterize the particular $f(R, T)$ model can also be written in terms of the cosmographic set, thus allowing such parameters to be restricted by the observational values of $(q_0, j_0, s_0)$ in order to tell us whether this specific form of $f(R, T)$ is a viable model or not.
	
	Recall that when obtaining the cosmography of the $f(R, T) = f(R) + f(T)$ model, we made the assumption that such a theory conserves the energy-momentum tensor, that is, $\nabla_{\mu}T^{\mu \nu}=0$. Although the non-conservation of the energy-momentum tensor is a remarkable feature of the $f(R, T)$ theories, there are cosmological and astrophysical reasons for considering the conservation of $T^{\mu \nu}$ in these theories, as one can see for example \cite{velten/2017, dos_santos/2019}. In \cite{alvarenga/2013, chakraborty/2013} it is shown that when we impose $\nabla_{\mu}T^{\mu \nu}=0$, $f(T)$ is restricted in such a way that its unique possible form for a universe filled with dust would be $f(T)=\alpha T^{1/2}$, where $\alpha$ is a parameter to be determined. We will consider below some particular models for which such an assumption is fulfilled.
	
	\subsubsection{$f(R, T)= R + \alpha T^{1/2}$ model}
	
	As a first case we will consider the $f(R, T)= R + \alpha T^{1/2}$ model. This form is particularly interesting because it carries the geometrical term that goes into the Einstein-Hilbert action of General Relativity \cite{de_felice/2010b}. For this model we have that at $t = t_0$
	
	\numparts
	\begin{eqnarray} \label{rd22} 
		f_0=R_0+ \alpha T_0^{1/2},\\
		f_{R}^{(0)}=1,\\ 
		f_{RR}^{(0)}=0,\\ 
		f_{RRR}^{(0)}=0.
	\end{eqnarray}
	\endnumparts
	The result $f_{R}^{(0)} = 1$ was already expected since we had previously imposed this value on the first derivative of $f(R, T)$ in (32). From these results we obtain that
	
	\begin{equation} \label{rd23}
		\alpha =\frac{f_0-R_0}{T_0^{1/2}}.
	\end{equation}
	So using the expressions of $f_0$ and $R_0$ in terms of cosmographic parameters, we can write $\alpha$ in terms of $ (q_0, j_0, s_0)$.
	
	\subsubsection{$f(R, T)=\alpha R^n + \beta T^{1/2}$ model}
	
	As a second case we will consider the $f(R, T)=\alpha R^n + \beta T^{1/2}$ model. The $f(R) = \alpha R^n$ form is interesting because it is a very general form and for certain values of the parameters $\alpha$ and $n$ we can retrieve the Einstein-Hilbert term. Moreover, it gives rise to an exact de Sitter solution needed to account for the inflationary paradigm \cite{de_felice/2010b}. For this model, $n$ is a positive real number, and $\alpha$ and $\beta$ are constants. When $t = t_0$ we get
	
	\numparts
	\begin{eqnarray} \label{rd23}
		f_0=\alpha R_0^n+ \beta T_0^{1/2},\\
		f_{R}^{(0)}=n \alpha R_0^{n-1},\\
		f_{RR}^{(0)}=n(n-1)\alpha R_0^{n-2},\\ f_{RRR}^{(0)}=n(n-1)(n-2)\alpha R_0^{n-3}.
	\end{eqnarray}
	\endnumparts
	From now on, for simplicity, we will use the following notation
	
	\numparts
	\begin{eqnarray} \label{rd24}
		f_0=\varphi_0,\\
		f_{R}^{(0)}=\varphi_1,\\
		f_{RR}^{(0)}=\varphi_2,\\ 
		f_{RRR}^{(0)}=\varphi_3.
	\end{eqnarray}
	\endnumparts
	We have then for the system of equations (76) that
	
	\numparts
	\begin{eqnarray} \label{rd25}
		\varphi_0=\alpha R_0^n+ \beta T_0^{1/2},\\
		\varphi_1=n\alpha R_0^{n-1},\\ 
		\varphi_2=n(n-1)\alpha R_0^{n-2},\\ 
		\varphi_3=n(n-1)(n-2)\alpha R_0^{n-3}. 
	\end{eqnarray}
	\endnumparts
	Solving this system of equations, we have 
	
	\begin{equation} \label{rd26}
		\beta= \frac{\varphi _0-\alpha R_0^n}{T_0^{1/2}}, 
	\end{equation}
	
	\begin{equation} \label{rd27}
		\alpha= \frac{\varphi _1 R_0^{1-n}}{n}, 
	\end{equation}
	
	\begin{equation} \label{rd28}
		n=\frac{\varphi_3R_0}{\varphi_2}+2.
	\end{equation}
	By using the expressions for $f_0$, $f_{R}^{(0)}$, $f_{RR}^{(0)}$, $f_{RRR}^{(0)}$ and $R_0$ in terms of cosmographic parameters, we can write $\alpha$, $\beta$ and $n$ in terms of $(q_0, j_0, s_0)$.
	
	\subsubsection{$f(R, T)=\alpha R + \beta R^n+ \gamma T^{1/2}$ model}
	
	As a last example, we will consider here the $f(R, T)=\alpha R + \beta R^n+ \gamma T^{1/2}$ model. The $f(R) = \alpha R + \beta R^n$ form is particularly interesting because for certain values assumed by the parameters $\alpha$, $\beta$ and $n$, we can retrieve both the Einstein-Hilbert term as the Starobinsky model, which so far is the most successful model of inflation \cite{de_felice/2010b, starobinsky/1980}. For this model, $n$ is a positive real number and $\alpha$, $\beta$ and $\gamma$ are constants. When $t = t_0$ we get
	
	\numparts
	\begin{eqnarray} \label{rd29} 
		f_0=\alpha R_0+ \beta R_0^n + \gamma T_0^{1/2},\\ 
		f_{R}^{(0)}=\alpha + n\beta R_0^{n-1},\\ 
		f_{RR}^{(0)}=n(n-1)\beta R_0^{n-2},\\ 
		f_{RRR}^{(0)}=n(n-1)(n-2)\beta R_0^{n-3}.
	\end{eqnarray}
	\endnumparts
	With the adopted notation,
	
	\numparts
	\begin{eqnarray} \label{rd30} 
		\varphi_0=\alpha R_0+ \beta R_0^n + \gamma T_0^{1/2},\\ 
		\varphi_1=\alpha + n\beta R_0^{n-1},\\ 
		\varphi_2=n(n-1)\beta R_0^{n-2},\\ 
		\varphi_3=n(n-1)(n-2)\beta R_0^{n-3}.
	\end{eqnarray}
	\endnumparts
	Solving this system of equations for the quantities of interest we obtain
	
	\begin{equation} \label{rd31}
		\gamma= \frac{\varphi_0- \alpha R_0- \beta R_0^n}{T_0^{1/2}},
	\end{equation}
	
	\begin{equation} \label{rd32}
		\alpha=\varphi_1-n\beta R_0^{n-1},
	\end{equation}
	
	\begin{equation} \label{rd33}
		\beta= \frac{\varphi_2R_0^{2-n}}{n(n-1)},
	\end{equation}
	
	\begin{equation} \label{rd34}
		n=\frac{\varphi_3R_0}{\varphi_2}+2.
	\end{equation}
	So, using the expressions for $f_0$, $f_{R}^{(0)}$, $f_{RR}^{(0)}$, $f_{RRR}^{(0)}$ and $R_0$ in terms of cosmographic parameters, we can write the parameters $\gamma$, $\alpha$, $\beta$ and $n$ in terms of $(q_0, j_0, s_0)$.
	
	\section{Conclusions} \label{sec:cc}
	
	Cosmography is a mathematical framework for the description of the universe based on the cosmological principle, being then inherently kinematic in the sense that it is independent of the dynamics obeyed by the scale factor $a(t)$. The most relevant property of cosmography is the capability of overcoming the so-called degeneracy problem among different cosmological models proposed to explain the current picture of the universe.
	
	Motivated by this possibility, we proposed here the cosmography of the conservative version of the $f(R, T)= f(R)+f(T)$ theory of gravity, which is given by Equations (\ref{cfrt18})-(\ref{cfrt20}).  A remarkable feature of the cosmographic equations is that they depend only on the parameters $(q_0, j_0, s_0)$. This is important in view of the fact that the higher order cosmographic parameters are poorly restricted by current cosmological data \cite{capozziello/2011c, aviles/2012, demianski/2012}. We have also shown here how these equations can be used together with the observational values of $(q_0, j_0, s_0)$ to place restrictions on the form of the $f(R, T)$ function in a similar way to what was done for the $f(R)$ theory \cite{capozziello/2008b, pannia/2013, aviles/2013}. In addition, the Equations (\ref{cfrt18})-(\ref{cfrt20}) have been rewritten within various dark energy scenarios. This can be useful due to the fact that as aforementioned the cosmographic approach can help to discriminate between different models of dark energy \cite{sahni/2003, alam/2003, rapetti/2007, demianski/2012}.
	
	Finally, it is worth mentioning that although cosmography is a robust tool to distinguish among competing cosmological models, it presents some difficulties. The limits of the standard cosmographic approach, based on the Taylor approximations, emerge when cosmological data at high redshifts like those from gamma-ray bursts are used \cite{izzo/2009, capozziello/2009, vincenzo/2010, xu/2011, demianski/2012, xia/2012}. In fact, the Taylor series converges if $z < 1$, so that any cosmographic analysis employing data disrespecting this limit is plagued by severe restrictions. Truncated approximations of the Taylor series are just approximations to an exact expression, what introduces errors into the analysis since the formulas used for fitting only represent approximations to the true expressions. This can be moderated by including higher orders of the series, but this comes at the expense of introducing more fitting parameters what in turn considerably complicates the corresponding statistical analysis. Thus, the convergence decreases when additional higher order terms are introduced, but on the other hand the accuracy of the analysis may be compromised by considering only lower orders. So, in order to avoid such problems, some modifications of the standard formalism have been proposed, among them we can mention the use of alternative redshift variables, which exhibit improved convergence properties as compared to the
	conventional redshift $z$ \cite{cattoen/2007, vincenzo/2010, capozziello/2020} and the use of rational approximations that aim to extend the radius of convergence of the Taylor series to high-redshift domains, amongst which the Pad\'e and Chebyshev polynomials represent  relevant examples \cite{gruber/2014, aviles/2014, capozziello/2018, capozziello/2020}.\\

	{\bf Acknowledgements}
	
	ISF and PHRSM thank CAPES for financial support.
	
	\newpage
	
	\flushleft
	

\textbf{References}
	
	\begin{thebibliography}{99}
		
		\bibitem{riess/1998} A.G. Riess et al., \textit{Astron. J.} {\bf 116}, 1009 (1998).
		\bibitem{perlmutter/1999} S. Perlmutter et al., \textit{Astrophys. J.} {\bf 517}. 565 (1999).
		\bibitem{farooq/2017} O. Farooq et al., \textit{Astrophys. J.} {\bf 835}, 26 (2017).
		\bibitem{wang/2006} F.Y. Wang and Z.G. Dai, \textit{Month. Not. Roy. Astron. Soc.} {\bf 368}, 371 (2006).
		\bibitem{cunha/2009} J.V. Cunha, \textit{Phys. Rev. D} {\bf 79}, 047301 (2009).
		\bibitem{ryden/2003} B. Ryden, {\it Introduction to Cosmology} (Addison Wesley, San Francisco, USA, 2003).
		\bibitem{guth/1981} A.H. Guth, \textit{Phys. Rev. D} {\bf 23}, 347 (1981).
		\bibitem{popa/2011} L.A. Popa, \textit{J. Cosm. Astrop. Phys.} {\bf 10}, 025 (2011).
		\bibitem{clutton/brock/1993} M. Clutton-Brock, \textit{Quart. J. Roy. Astron. Soc.} {\bf 34}, 411 (1993).
		\bibitem{shtanov/2010} Y. Shtanov, \textit{Ann. Phys.} {\bf 19}, 332 (2010).
		\bibitem{guzzetti/2016} M.C. Guzzetti et al., \textit{Riv. Nuo. Cim.} {\bf 39}, 399 (2016).  
		\bibitem{weinberg/1989} S. Weinberg, \textit{Rev. Mod. Phys.} {\bf 61}, 1 (1989).
		\bibitem{nesseris/2006} S. Nesseris and L. Perivolaropoulos, \textit{Phys. Rev. D} {\bf 73}, 103511 (2006).
		\bibitem{chiba/2003} T. Chiba, \textit{Phys. Lett. B} {\bf 575}, 1 (2003).
		\bibitem{capozziello/1998} S. Capozziello et al., \textit{Gen. Rel. Grav.} {\bf 30}, 1247 (1998).
		\bibitem{capozziello/2011} S. Capozziello and M. de Laurentis, \textit{Phys. Rep.} {\bf 509}, 167 (2011).
		\bibitem{capozziello/2008} S. Capozziello and M. Francaviglia, \textit{Gen. Rel. Grav.} {\bf 40}, 357 (2008).
		\bibitem{suvorov/2019} A.G. Suvorov, \textit{Phys. Rev. D} {\bf 99}, 124026 (2019).
		\bibitem{cognola/2005} G. Cognola et al., \textit{J. Cosm. Astrop. Phys.} {\bf 02}, 010 (2005).
		\bibitem{azadi/2008} A. Azadi et al., \textit{Phys. Lett. B} {\bf 670}, 210 (2008).
		\bibitem{olmo/2015} G. Olmo and D. Rubiera-Garcia, \textit{Universe} {\bf 1}, 173 (2015).
		\bibitem{momeni/2009} D. Momeni and H. Gholizade, \textit{Int. J. Mod. Phys. D} {\bf 18}, 1719 (2009).
		\bibitem{santos/2007} J. Santos et al., \textit{Phys. Rev. D} {\bf 76}, 083513 (2007).
		\bibitem{santos/2010} J. Santos et al., \textit{Int. J. Mod. Phys. D} {\bf 19}, 1315 (2010). 
		\bibitem{harko/2010} T. Harko and F.S.N. Lobo, \textit{Eur. Phys. J. C} {\bf 70}, 373 (2010).
		\bibitem{azevedo/2016} R.P.L. Azevedo and J. P\'aramos, \textit{hys. Rev. D} {\bf 94}, 064036 (2016). 
		\bibitem{harko/2011} T. Harko et al., \textit{Phys. Rev. D} \textbf{84}, 024020 (2011).
		\bibitem{alves/2016} M.E.S. Alves, P.H.R.S. Moraes, J.C.N. de Araujo and M. Malheiro, \textit{Phys. Rev. D} {\bf 94}, 024032 (2016).
		\bibitem{moraes/2016} P.H.R.S. Moraes et al., \textit{J. Cosm. Astrop. Phys.} {\bf 6}, 005 (2016).
		\bibitem{alvarenga/2013} F.G. Alvarenga et al., \textit{Phys. Rev. D} {\bf 87}, 103526 (2013).
		\bibitem{reddy/2012} D.R.K. Reddy et al., \textit{Astrophys. Spa. Sci.} {\bf 342}, 249 (2012).
		\bibitem{shabani/2013} H. Shabani and M. Farhoudi, \textit{Phys. Rev. D} {\bf 88}, 044048 (2013). 
		\bibitem{giribet/2006} G. Giribet et al., \textit{J. High Ener. Phys.} {\bf 05}, 007 (2006).
		\bibitem{dominguez/2006} A.E. Dominguez and E. Gallo, \textit{Phys. Rev. D} {\bf 73}, 064018 (2006).
		\bibitem{torii/1998} T. Torii and K.-I. Maeda, \textit{Phys. Rev. D} {\bf 58}, 084004 (1998).
		\bibitem{chen/2007} C.-M. Chen et al., \textit{Phys. Rev. D} {\bf 75}, 084030 (2007).
		\bibitem{myrzakulov/2011} R. Myrzakulov et al., \textit{Gen. Rel. Grav.} {\bf 43}, 1671 (2011).
		\bibitem{zhou/2009} S.-Y. Zhou et al., \textit{J. Cosm. Astrop. Phys.} {\bf 07}, 009 (2009).
		\bibitem{de_felice/2009b} A. de Felice and S. Tsujikawa, \textit{Phys. Lett. B} {\bf 675}, 1 (2009).
		\bibitem{de_felice/2009} A. de Felice and S. Tsujikawa, \textit{Phys. Rev. D} {\bf 80}, 063516 (2009).
		%\bibitem{de_felice/2010} A. de Felice and T. Tanaka, Prog. Theor. Phys. {\bf 124}, 503 (2010).
		%\bibitem{de_felice/2011} A. de Felice et al., Phys. Rev. D {\bf 83}, 104035 (2011).
		\bibitem{de_felice/2010} A. de Felice et al., \textit{Phys. Rev. D} {\bf 82}, 063526 (2010).
		\bibitem{elizalde/2010} E. Elizalde et al., \textit{Class. Quant. Grav.} {\bf 27}, 095007 (2010).
		\bibitem{de_la_cruz-dombriz/2012} \'A. de la Cruz-Dombriz and D. S\'aez-G\'omez, \textit{Class. Quant. Grav.} {\bf 29}, 245014 (2012).
		\bibitem{de_laurentis/2015} M. de Laurentis et al., \textit{Phys. Rev. D} {\bf 91}, 083531 (2015).
		\bibitem{santos_da_costa/2018} S. Santos da Costa et al., \textit{Class. Quant. Grav.} {\bf 35}, 075013 (2018).
		\bibitem{cai/2016} Y.-F. Cai et al., \textit{Rep. Prog. Phys.} {\bf 79}, 106901 (2016).
		\bibitem{ferraro/2007} R. Ferraro and F. Fiorini, \textit{Phys. Rev. D} {\bf 75}, 084031 (2007).
		\bibitem{obukhov/2003} Y.N. Obukhov and J.G. Pereira, \textit{Phys. Rev. D} {\bf 67}, 044016 (2003).
		\bibitem{bamba/2013} K. Bamba et al., \textit{Phys. Rev. D} {\bf 88}, 084042 (2013).
		\bibitem{sotiriou/2011} T.P. Sotiriou et al., \textit{Phys. Rev. D} {\bf 83}, 104030 (2011).
		\bibitem{golovnev/2017} A. Golovnev et al., \textit{Class. Quant. Grav.} {\bf 34}, 145013 (2017).
		\bibitem{vargas/2004} T. Vargas, \textit{Gen. Rel. Grav.} {\bf 36}, 1255 (2004).
		\bibitem{de_andrade/2000} V.C. de Andrade et al., \textit{Phys. Rev. Lett.} {\bf 84}, 4533 (2000).
		\bibitem{harko/2014} T. Harko et al., \textit{J. Cosm. Astrop. Phys.} {\bf 12}, 021 (2014).
		\bibitem{chattopadhyay/2014} S. Chattopadhyay et al., \textit{Astrophys. Spa. Sci.} {\bf 353}, 279 (2014).
		\bibitem{kofinas/2014} G. Kofinas et al., \textit{Class. Quant. Grav.} {\bf 31}, 175011 (2014).
		\bibitem{kofinas/2014b} G. Kofinas and E.N. Saridakis, \textit{Phys. Rev. D} {\bf 90}, 084045 (2014).
		\bibitem{sharif/2016} M. Sharif and A. Ikram, \textit{Eur. Phys. J. C} {\bf 76}, 640 (2016).
		\bibitem{sharif/2018} M. Sharif and A. Ikram, \textit{Astrophys. Spa. Sci.} {\bf 363}, 178 (2018).
		\bibitem{bhatti/2018} M.Z.-U.-H., Bhatti et al., \textit{Int. J. Mod. Phys. D} {\bf 27}, 1850044 (2018).
		\bibitem{yousaf/2018} Z. Yousaf, \textit{Astrophys. Spa. Sci.} {\bf 363}, 226 (2018).
		%\bibitem{shamir/2017} M.F. Shamir and M. Ahmad, Eur. Phys. J. C {\bf 77}, 55 (2017).
		%\bibitem{shamir/2018} M.F. Shamir and M. Ahmad, Phys. Rev. D {\bf 97}, 104031 (2018).
		\bibitem{mak/2002} M.K. Mak and T. Harko, \textit{Int. J. Mod. Phys. D} {\bf 11}, 1389 (2002).
		\bibitem{amendola/2000} L. Amendola, \textit{Phys. Rev. D} {\bf 62}, 043511 (2000).
		\bibitem{huey/2006} G. Huey and B.D. Wandelt, \textit{Phys. Rev. D} {\bf 74}, 023519 (2006).
		\bibitem{sahni/2002} V. Sahni, \textit{Class. Quant. Grav.} {\bf 19}, 3435 (2002).
		\bibitem{linder/2008} E.V. Linder, \textit{Gen. Rel. Grav.} {\bf 40}, 329 (2008).
		\bibitem{bento/2003} M.C. Bento et al., \textit{Phys. Lett. B} {\bf 575}, 172 (2003).
		\bibitem{bento/2003b} M.C. Bento et al., \textit{Gen. Rel. Grav.} {\bf 35}, 2063 (2003)
		\bibitem{chakraborty/2007} W. Chakraborty et al., \textit{Grav. Cosm.} {\bf 13}, 293 (2007).
		\bibitem{moraes/2016b} P.H.R.S. Moraes, \textit{Int. J. Mod. D} {\bf 25}, 1650009 (2016). 
		\bibitem{trodden/2007} M. Trodden, \textit{Int. J. Mod. Phys. D} {\bf 16}, 2065 (2007).
		\bibitem{trodden/2011} M. Trodden, \textit{Gen. Rel. Grav.} {\bf 43}, 3367 (2011).
		\bibitem{tsujikawa/2010} S. Tsujikawa, \textit{Lect. Not. Phys.} {\bf 800}, 99 (2010).
		\bibitem{tsujikawa/2013} S. Tsujikawa, \textit{Class. Quant. Grav.} {\bf 30}, 214003 (2013).
		\bibitem{bloomfield/2013} J. Bloomfield et al., \textit{J. Cosm. Astrop. Phys.} {\bf 08}, 010 (2013).
		\bibitem{chimento/2000} L.P. Chimento et al., \textit{Phys. Rev. D} {\bf 62}, 063508 (2000).
		\bibitem{olivares/2008} G. Olivares et al., \textit{Phys. Rev. D} {\bf 77}, 063513 (2008).
		\bibitem{bamba/2012} K. Bamba et al., \textit{Astrophys. Spa. Sci.} {\bf 342}, 155 (2012).
		\bibitem{yang/2019} W. Yang et al., \textit{Phys. Rev. D} {\bf 100}, 023522 (2019).
		\bibitem{weinberg/1972} S. Weinberg, {\it Gravitation and Cosmology} (Wiley, New York, USA, 1972).
		\bibitem{visser/2004} M. Visser, \textit{Class. Quant. Grav.} {\bf 21}, 2603 (2004).
		\bibitem{visser/2005} M. Visser, \textit{Gen. Rel. Grav.} {\bf 37}, 1541 (2005).
		\bibitem{pannia/2013} F.A.T. Pannia and S.E.P. Bergliaffa, \textit{J. Cosm. Astrop. Phys.} {\bf 08}, 030 (2013).
		\bibitem{pizza/2015} L. Pizza, \textit{Phys. Rev. D} {\bf 91}, 124048 (2015).
		\bibitem{pires/2010} N. Pires et al., \textit{Phys. Rev. D} {\bf 82}, 067302 (2010).
		\bibitem{aviles/2013} A. Aviles et al., \textit{Phys. Rev. D} {\bf 87}, 044012 (2013).
		\bibitem{bouhmadi-lopez/2010} M. Bouhmadi-L\'opez et al., \textit{Phys. Rev. D} {\bf 82}, 103526 (2010).
		\bibitem{capozziello/2008b} S. Capozziello et al., \textit{Phys. Rev. D} {\bf 78}, 063504 (2008).
		\bibitem{capozziello/2011b} S. Capozziello et al., \textit{Phys. Rev. D} {\bf 84}, 043527 (2011).
		\bibitem{piedipalumbo/2015} E. Piedipalumbo et al., \textit{Int. J. Mod. Phys. D} {\bf 24}, 1550100 (2015).
		\bibitem{capozziello/2019} S. Capozziello et al., \textit{Int. J. Mod. Phys. D} {\bf 28}, 1930016 (2019). 
		\bibitem{barrientos/2018} E. Barrientos et al., \textit{Phys. Rev. D} {\bf 97}, 104041 (2018).
		\bibitem{vincenzo/2010} V. Vitagliano et al., \textit{J. Cosm. Astrop. Phys.} {\bf 03}, 005 (2010)
		\bibitem{cai/2016b} R.-G. Cai et al., \textit{Phys. Rev. D} {\bf 93}, 043517 (2016).
		\bibitem{ruderman/1972} M. Ruderman, \textit{Ann. Rev. Astron. Astrophys.} {\bf 10}, 427 (1972).
		\bibitem{canuto/1974} V. Canuto and S.M. Chitre, \textit{Phys. Rev. D} {\bf 8}, 1587 (1974).
		\bibitem{moraes/2018} P.H.R.S. Moraes et al., \textit{Eur. Phys. J. C} {\bf 78}, 192 (2018).
		\bibitem{geng/2017} C.-Q. Geng et al., \textit{J. Cosm. Astrop. Phys.} {\bf 08}, 032 (2017).
		\bibitem{singh/2016} V. Singh and C.P. Singh, \textit{Int. J. Theor. Phys.} {\bf 55}, 1257 (2016).
		\bibitem{kumar/2015} P. Kumar and C.P. Singh, \textit{Astrophys. Spa. Sci.} {\bf 357}, 120 (2015).
		\bibitem{harko/2014b} T. Harko, \textit{Phys. Rev. D} {\bf 90}, 044067 (2014).
		\bibitem{asadiyan/2019} K. Asadiyan et al., \textit{Int. J. Geom. Meth. Mod. Phys.} {\bf 16}, 1950153 (2019).
		\bibitem{dos_santos/2019} S.I. dos Santos, G.A. Carvalho, P.H.R.S. Moraes, C.H. Lenzi and M. Malheiro, \textit{Eur. Phys. J. Plus} {\bf 134}, 398 (2019).
		\bibitem{carvalho/2020} G.A. Carvalho, S.I. dos Santos, P.H.R.S. Moraes and M. Malheiro, \textit{Int. J. Mod. Phys. D} {\bf 29}, 2050075 (2020).
		\bibitem{alvarenga/2013} F.G. Alvarenga et al., \textit{Phys. Rev. D} {\bf 87}, 103526 (2013).
		\bibitem{chakraborty/2013} S. Chakraborty, \textit{Gen. Rel. Grav.} {\bf 45}, 2039 (2013). 
		\bibitem{velten/2017} H. Velten and T.R.P. Caramês, \textit{Phys. Rev. D} {\bf 95}, 123536 (2017)
		\bibitem{capozziello/2005} S. Capozziello et al., \textit{Phys. Rev. D} {\bf 71}, 043503 (2005)
		\bibitem{barrientos/2014} J. Barrientos and G. F. Rubilar, \textit{Phys. Rev. D} {\bf 90}, 028501 (2014)
		\bibitem{moraes/2017} P.H.R.S Moraes and P.K. Sahoo, \textit{Eur. Phys. J. C} {\bf 77}, 480 (2017)
		\bibitem{moraes/2016c} P.H.R.S Moares and J.R.L Santos, \textit{Eur. Phys. J. C} {\bf 76}, 60 (2016)
		\bibitem{moraes/2016d} P.H.R.S Moraes and R.A.C Correa, \textit{Astrophys. Spa. Sci.} {\bf 361}, 227 (2016)
		\bibitem{taskforce} A. Albrecht et al., \textit{Fermilab Report No. FERMILAB-FN-0793-A}, 2006.
		\bibitem{chevallier} M. Chevallier and D. Polarski, \textit{Int. J. Mod. Phys. D}, {\bf 10}, 2 (2001)
		\bibitem{linder/2003} E. Linder, \textit{Phys. Rev. Lett.} {\bf 90}, 9 (2003)
		\bibitem{capozziello/2011c} S. Capozziello et al., \textit{Phys. Rev. D} {\bf 84}, 124061 (2011)
		%\bibitem{velten/2017} H. Velten and T.R.P. Caramês, Phys. Rev. D {\bf 95}, 123536 (2017)
		\bibitem{de_felice/2010b} A. De Felice and S. Tsujikawa, \textit{Living Rev. Relativity}, {\bf 13}, 3 (2010)
		\bibitem{starobinsky/1980} A.A. Starobinsky, \textit{Phys. Lett. B} {\bf 91}, 1 (1980)
		\bibitem{aviles/2012} A. Aviles, C. Gruber, O. Luongo and H. Quevedo, \textit{Phys. Rev. D} {\bf 86}, 123516 (2012)
		\bibitem{demianski/2012} M. Demianski, E. Piedipalumbo, C. Rubano and P. Scudellaro, \textit{Mon. Not. R. Astron. Soc.} {\bf 426}, 1396 (2012)
		\bibitem{rapetti/2007} D. Rapetti, S.W. Allen, M.A. Amin and R.D. Blandford, \textit{Mon. Not. R. Astron. Soc.} {\bf 375}, 1510–1520 (2007)
		\bibitem{sahni/2003} V. Sahni, T.D. Saini, A.A. Starobinsky and U. Alam, \textit{JETP Lett.} {\bf 77}, 5 (2003)
		\bibitem{alam/2003} U. Alam, V. Sahni, T.D. Saini and A.A. Starobinsky, \textit{Mon. Not. R. Astron. Soc.} {\bf 344}, 1057 (2003)
		\bibitem{xia/2012} J. Xia, V. Vitagliano, S. Liberati and Matteo Viel, \textit{Phys. Rev. D} {\bf 85}, 043520 (2012)
		\bibitem{xu/2011} L. Xu and Y. Wang, \textit{Phys. Lett.} {\bf 702}, 114 (2011)
		\bibitem{izzo/2009} L. Izzo,  S. Capozziello, G. Covone and M. Capaccioli, \textit{Astron. \& Astrop.} {\bf 508}, 67 (2009)
		\bibitem{capozziello/2009} S. Capozziello and L. Izzo,  \textit{Nuc. Phys. B} {\bf 194}, 206 (2009)
		\bibitem{cattoen/2007} C. Catt\"{o}en and M. Visser, Class. \textit{Quantum Grav.} {\bf 24}, 5985 (2007)
		\bibitem{capozziello/2020} S. Capozziello, R. D’Agostino  and O. Luongo, \textit{Mon. Not. R. Astron. Soc.} {\bf 494}, 2576 (2020)
		\bibitem{gruber/2014} C. Gruber and O. Luongo, \textit{Phys. Rev. D} {\bf 89}, 103506 (2014)
		\bibitem{aviles/2014} A. Aviles, A. Bravetti, S. Capozziello and O. Luongo, \textit{Phys. Rev. D} {\bf 90}, 043531 (2014)
		\bibitem{capozziello/2018} S. Capozziello, R. D’Agostino  and O. Luongo,  \textit{Mon. Not. R. Astron. Soc.} {\bf 476}, 3924 (2018)
		%\bibitem{xia/2012} J. Xia, V. Vitagliano, S. Liberati and Matteo Viel, Phys. Rev. D {\bf 85}, 043520 (2012)
		%\bibitem{xu/2011} L. Xu and Y. Wang, Phys. Lett. {\bf 702}, 114 (2011)
		%\bibitem{izzo/2009} L. Izzo,  S. Capozziello, G. Covone and M. Capaccioli, Astron. \& Astrop. {\bf 508}, 67 (2009)
		%\bibitem{capozziello/2009} S. Capozziello and L. Izzo,  Nuc. Phys. B {\bf 194}, 206 (2009)
		
	\end{thebibliography}
\end{document}